\documentclass{emulateapj}
\shorttitle{Planets on wide orbits}
\shortauthors{Dodson-Robinson et al.}
\usepackage{multirow}
\begin{document}

\title{The Formation Mechanism of Gas Giants on Wide Orbits}

\author{Sarah E. Dodson-Robinson\altaffilmark{1,4},
Dimitri Veras\altaffilmark{2},
Eric B. Ford\altaffilmark{2},
C. A. Beichman\altaffilmark{3}}

\altaffiltext{1}{Astronomy Department, University of Texas, 1 University
Station C1400, Austin, TX 78712; sdr@astro.as.utexas.edu}

\altaffiltext{2}{Astronomy Department, University of Florida, 211 Bryant
Space Sciences Center, Gainesville, FL 32111, USA}

\altaffiltext{3}{NASA Exoplanet Science Institute, California Institute
of Technology, 770 S. Wilson Ave, Pasadena, CA 91125, USA}

\altaffiltext{4}{Formerly Sarah E. Robinson}

\begin{abstract}
The recent discoveries of massive planets on ultra-wide orbits orbiting
HR 8799 (Marois et al. 2008) and Fomalhaut (Kalas et al. 2008) present a
new challenge for planet formation theorists. Our goal is to figure out
which of three giant planet formation mechanisms---core accretion (with
or without migration), scattering from the inner disk, or gravitational
instability---could be responsible for Fomalhaut b, HR 8799 b, c and d,
and similar planets discovered in the future. This paper presents the
results of numerical experiments comparing the long-period planet
formation efficiency of each possible mechanism in model A star, G star
and M star disks.

First, a simple core accretion simulation shows that planet cores
forming beyond 35~AU cannot reach critical mass, even under the most
favorable conditions one can construct. Second, a set of N-body
simulations demonstrates that planet-planet scattering does not create
stable, wide-orbit systems such as HR~8799. Finally, a linear stability
analysis verifies previous work showing that global spiral instabilities
naturally arise in high-mass disks. {\it We conclude that massive gas
giants on stable orbits with semimajor axes $a \ga 35$~AU form by
gravitational instability in the disk.} We recommend that observers
examine the planet detection rate as a function of stellar age,
controlling for the planets' dimming with time. Any age trend would
indicate that planets on wide orbits are transient relics of scattering
from the inner disk. If planet detection rate is found to be independent
of stellar age, it would confirm our prediction that gravitational
instability is the dominant mode of producing detectable planets on wide
orbits. We also predict that the occurrence ratio of long-period to
short-period gas giants should be highest for M dwarfs due to the
inefficiency of core accretion and the expected small fragment mass
($\sim 10 M_{\rm Jup}$) in their disks.

\end{abstract}

\keywords{planetary systems --- planetary systems: formation ---
accretion disks --- stars: formation --- instabilities}

\section{Introduction}
\label{intro}

The first direct images of extrasolar gas giants present planetary
scientists with a new mandate: construct massive planets on extremely
wide orbits. Before November 2008, when the existence of planets
surrounding Fomalhaut (Kalas et al.\ 2008) and HR~8799 (Marois et al.\
2008) was revealed, theorists had reached a near-consensus that the core
accretion mechanism could explain nearly all observed properties of
planets discovered by radial velocities. Simple core accretion-based
models reproduce the planet-metallicity correlation (Fischer \& Valenti
2005) and the planet-silicon correlation (Robinson et al.\ 2006), and
the rarity of giant planets orbiting M dwarfs (Laughlin et al.\ 2004;
Johnson et al.\ 2007).  Unfortunately, with the exception of Ida \& Lin
(2004), these models treat only planets forming near the ice line
and ignore planet formation beyond $\sim 5$~AU---for the simple reason
that the vast majority of known radial-velocity planets orbit within
3~AU of their host stars.  Estimates of the planet occurrence rate
beyond 5~AU (e.g.\ Cumming et al.\ 2008) rely on extrapolation of the
statistics on shorter-period orbits.

One can immediately foresee problems applying the core accretion
model---in which a massive solid core destabilizes the surrounding disk
gas to accrete an atmosphere---to Fomalhaut~b, located 119~AU from the
host star, and HR~8799~b, c and d, with projected stellar separations of
24, 38 and 68~AU. All of these new planets are unequivocally massive and
gaseous: the smallest is Fomalhaut~b with $M < 3 M_{\rm Jup}$. Even
forming the small and solid-rich Uranus and Neptune (14.2 and 17.5
$M_{\oplus}$, respectively) {\it in situ} is difficult: Levison \&
Stewart (2001) show that the planets cannot form in their current orbits
at 19 and 30~AU because the accretion timescale is far longer than the
planetesimal excitation timescale. If accretion timescale prohibits
medium-mass planet formation at 19~AU, the formation of a gas giant at
$> 100$~AU by core accretion hardly seems possible.

There are a few ways to increase the efficiency of core accretion.
Perhaps the protoplanetary disks surrounding HR~8799 and Fomalhaut held
far higher planetesimal masses than the solar nebula, and were thus
better able to form solid cores at large radii. Since both stars host
massive, bright debris belts (Kalas et al.\ 2005; Chen et al. 2009),
this hypothesis requires investigation. It's also possible that these
relatively massive stars ($\sim 1.5 M_{\odot}$, $\sim 2.0 M_{\odot}$
respectively) decreased the protoplanet orbital timescale in the outer
disk enough for core accretion to proceed. Perhaps it is not a
coincidence that the only directly imaged extrasolar planets orbit A
stars. In this work we use model A, G and M-star disks to test the
possibility that core accretion at large radii could work for HR~8799
and Fomalhaut where it fails for lower-mass stars.

What are the other possible formation mechanisms for gas giants on wide
orbits? For truly massive planets---and in the case of HR~8799, the
best-fit masses of 7~$M_{\rm Jup}$, 10~$M_{\rm Jup}$ and 10~$M_{\rm
Jup}$ are near the Y dwarf minimum mass (Zuckerman \& Song 2009)---the
simplest possibility is fragmentation of the collapsing protostellar
core, analogous to binary star formation (e.g. Cha \& Whitworth 2003).
However, several lines of observational evidence show that HR~8799 and
Fomalhaut are not simply binary or multiple stars with high mass ratios:

\begin{enumerate}

\item Resolved images show that both stars host debris disks that are
confined by the planet(s), indicating that the planets and debris disks
are coplanar relics of protostellar disks (Kalas et al.\ 2005; Chen et
al.\ 2009; Su et al.\ 2009).

\item Spectroastrometric measurements of the spin-orbit alignment of
Fomalhaut show that the star's rotation axis is perpendicular to the
debris disk and planet orbit, further indicating that Fomalhaut b formed
in a disk (Le Bouquin et al.\ 2009). The rotational velocity of HR~8799
indicates that it, too, is nearly pole-on and perpendicular to the
planet orbits (Reidemeister et al.\ 2009).

\item The three companions to HR~8799 orbit the central star, rather
than forming a double-double system consistent with cloud core
fragmentation (Clark \& Bonnell 2005).

\end{enumerate}

As molecular cloud core fragmentation is unlikely, we will treat the
companions to HR~8799 and Fomalhaut as bona fide planets (whether they
burn deuterium or not) and search for formation mechanisms that take
place in protoplanetary disks. The two remaining possibilities are
planet-planet scattering, in which planets form in the inner disk and
are ejected to wide orbits by a yet-unseen inner planet, and
gravitational instability, where the protostellar disk fragments into
bound clumps that eventually contract to near Jupiter's radius. Here,
too, we run into problems. Planet scattering tends to create either
unstable systems or extremely eccentric orbits (Raymond et al.\ 2008;
Veras et al.\ 2009), whereas the orbits of HR~8799 b, c and d and
Fomalhaut b appear to be nearly circular. There is considerable debate
in the literature over whether disk fragments formed in gravitational
instabilities survive to become protoplanets or are destroyed by
accretion onto the star or background shear flows (Durisen et al.\ 2007;
Vorobyov \& Basu 2006); furthermore, a previous investigation of planet
formation on wide orbits by Boss (2006) noted that spiral instabilities
may wind too tightly at large radii for the disk to fragment at all.
Nonetheless, two points in favor of gravitational instability are that
spiral waves---whether they fragment the disk or not---develop most
easily in the outer parts disks, and that when fragments do form they
are often near $10 M_{\rm Jup}$ (Boley 2009; Stamatellos \& Whitworth
2009).  Gravitational instability is therefore perfectly positioned to
create massive planets with extremely long-period orbits.

In this paper we present numerical experiments that test the viability
of core accretion, planet-planet scattering and gravitational
instability. The paper is organized as follows. In \S \ref{diskmodel},
we describe the disk models that form the foundation of our numerical
experiments. In \S \ref{coreaccretion}, we use a simple core-accretion
simulation to show that massive core growth on ultra-wide orbits is not
possible. We present N-body simulations assessing the likelihood of
scattered planets landing in stable, wide orbits in \S \ref{scattering}.
The disk stability analysis in \S \ref{gi} confirms that our
maximum-mass model disks can become unstable to global two-armed spiral
modes. Finally, we present our conclusions, along with a proposed
observational experiment to confirm that gravitational instability is
the dominant mode of planet formation on wide orbits, in \S
\ref{conclusions}.

\section{Disk Model}
\label{diskmodel}

We begin by describing the simple disk model that forms the basis for
our three numerical experiments. In order to facilitate comparisons of
planet formation efficiency among different types of stars, we have
deliberately chosen a simple, passively illuminated disk whose
temperature profile depends only on the temperature and luminosity of
the host protostar.

We use the method of Chiang \& Goldreich (1997) to construct flared
protostellar disks that are in radiative equilibrium with their host
protostars. In the Chiang \& Goldreich models, dust in a superheated
surface layer reprocesses the incident stellar radiation, re-radiating
$\sim 1/2$ of the absorbed flux onto the disk interior at infrared
wavelengths. The midplane temperature of a disk optically thick to
visible stellar radiation is
\begin{equation}
T_i \approx \left ( \frac{\alpha}{4} \right )^{1/4} \left (
\frac{R_*}{a} \right )^{1/2} T_* .
\label{tint}
\end{equation}
In Equation \ref{tint}, $a$ is the distance from the star, $R_*$ is the
protostar radius, $T_*$ is the effective temperature and $\alpha$ is the
disk flaring angle:
\begin{equation}
\alpha = \frac{0.4 R_*}{a} + a \frac{d}{da} \left ( \frac{H}{a} \right
) .
\label{flaring}
\end{equation}
The first term in Equation \ref{flaring} can be neglected in the limit
$a \gg R_*$.

The visible photosphere height is given by
\begin{equation}
\frac{H}{a} = \frac{H}{h} \left ( \frac{T_*}{T_c} \right )^{4/7} \left (
\frac{a}{R_*} \right )^{2/7} ,
\label{Hoa}
\end{equation}
where $h$ is the pressure scale height. The quantity $T_c$ defines the
escape temperature of the disk gas from the stellar surface:
\begin{equation}
T_c = \frac{G M_* \mu}{k R_*} .
\label{tc}
\end{equation}
In Equation \ref{tc}, $M_*$ is the star mass, $\mu$ is the mean
molecular weight of the nebula gas in grams and $k$ is Boltzmann's
constant. We take $H/h = 4$ as a constant among all model disks.
Best-fit SEDs of protostellar disks have values $H/h$ between 2 and 5
(Chiang et al.\ 2001).

\begin{figure}
\epsscale{0.8}
\plotone{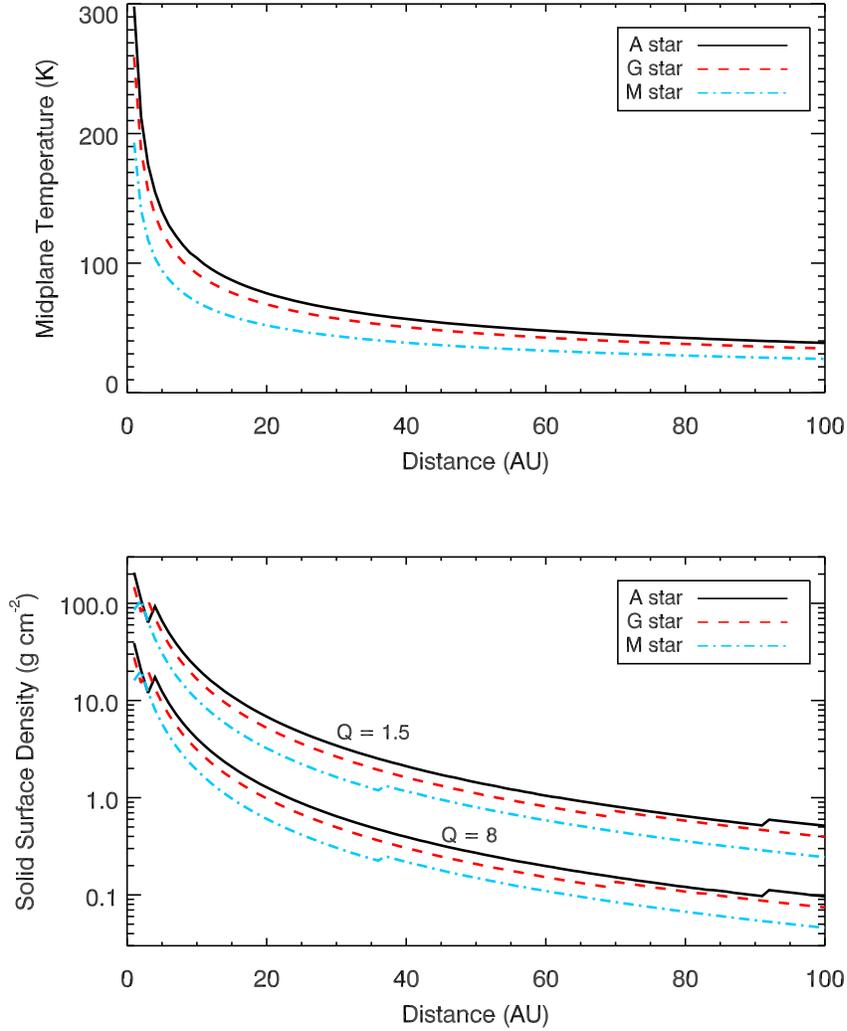}
\caption{A passive disk in radiative equilibrium with its parent star
forms the basis of our numerical experiments. {\bf Top}: midplane
temperature as a function of distance from the star for our three model
protostars, the $1.5 M_{\odot}$ A star, the $1 M_{\odot}$ G star and the
$0.5 M_{\odot}$ M star. {\bf Bottom}: Solid surface density as a
function of distance. The two sets of curves represent the maximum-mass
nebulae, with Toomre $Q = 1.5$, and the medium-mass nebulae, with $Q =
8$. The two discontinuities in the surface density distribution are the
water ice line ($< 5$ AU) and the methane condensation front. As long as
the nebula is optically thick, midplane temperature is independent of
solid surface density.}
\label{diskfig}
\end{figure}

We test the planet formation efficiency in disks surrounding the types
of stars known to host planets (spectral types A-M) by defining three
representative protostars: an A star of mass $1.5 M_{\odot}$, a G star
of mass $1.0 M_{\odot}$, and an M star of mass $0.5 M_{\odot}$. The
remaining inputs to Equations \ref{tint}--\ref{tc}, the star radius and
effective temperature, are taken from published pre-main-sequence
evolutionary tracks (D'Antona \& Mazzitelli 1994). We choose a stellar
age of $10^5$~years as the onset of planet formation. Our stars are
older than the 100~km planetesimal formation timescale (Barnes et al.\
2009) yet still young enough to harbor massive disks, which improves the
chances of planet formation. Figure \ref{diskfig} (top panel) shows the
midplane temperature as a function of radius for each protostar mass.

We must now calculate the mass available for planet formation. The
Chiang \& Goldreich temperature profile is independent of the disk
surface density as long as the disk is optically thick. We therefore use
the constraint of local gravitational stability to set the surface
density profile. The Toomre Q parameter, \begin{equation} Q = \frac{c_s
\kappa}{\pi G \Sigma} , \label{toomreq} \end{equation} measures the
stability of the disk to axisymmetric perturbations. In Equation
\ref{toomreq}, $\kappa$ is the epicyclic frequency, $c_s$ is the
midplane sound speed and $\Sigma$ is the gas surface density. Rings of
enhanced surface density, with corresponding rarefactions, will grow
exponentially in any part of the disk with $Q < 1$. With $c_s$
determined by the radiative equilibrium model and the assumption of a
nearly Keplerian disk, we can recover the surface density profile from
Equations \ref{tint}--\ref{toomreq} by specifying a value of $Q$:
\begin{equation}
\Sigma = \frac{1}{Q \pi} \left ( \frac{M_*}{G} \frac{k}{\mu} \right
)^{1/2} \left ( \frac{1}{14} \frac{H}{h} \right )^{1/8} T_*^{4/7} \:
T_c^{-1/14} \: R_*^{3/14} \: a^{-12/7} .
\label{sdensprofile}
\end{equation}

Since planets on wide orbits are known to be difficult to build (Kenyon
\& Bromley 2008; Whitworth \& Stamatellos 2006), we construct our disks
with the following question in mind: Are there {\it any} circumstances,
however optimistic, under which massive planets between 20 and 120~AU
can form and survive? We therefore outfit each star with a
``maximum-mass nebula'', a disk that contains as much mass as it can
plausibly hold while maintaining the disklike structure that is critical
for core accretion. The maximum-mass nebulae have $Q = 1.5$ everywhere
in the disk.

Furthermore, since most known planets are not on wide orbits and do not
require extremely massive disks to form, we also construct ``medium-mass
nebulae'' with $Q = 8$ for each star. The medium-mass nebulae allow us
to verify that our calculations of planet formation efficiency are
consistent with previous studies of both core accretion and
gravitational instability (e.g. Laughlin et al.\ 2004; Kenyon \& Bromley
2008; Kratter et al.\ 2008; Stamatellos \& Whitworth 2009).

The final step in constructing our disk models is to account for the
viscous accretion that happens before the onset of planet formation,
which tends to flatten a strongly centrally peaked surface density
profile (Lynden-Bell \& Pringle 1974; Dodson-Robinson et al.\ 2009). We
calculate the viscous diffusion timescale, over which surface density
decreases by a factor of $\sim 2$, as
\begin{equation}
t_{\rm diff} = \frac{a^2}{\nu} = \frac{a^2}{(\alpha_{\rm visc} c_s^2 /
\Omega)^2} ,
\label{tdiff}
\end{equation}
where $\alpha_{\rm visc}$ is the viscous efficiency in the standard
$\alpha$-disk model (Shakura \& Syunyaev 1973). Following
Dodson-Robinson et al.\ (2009) and Lyra et al.\ (2008), we take
$\alpha_{\rm visc} = 0.002$. Our final surface density is determined by
reducing the fiducial value according to
\begin{equation}
\Sigma = \Sigma_0 \left ( 2^{-10^5 / t_{\rm diff}} \right ),
\label{sigdiff}
\end{equation}
where $t_{\rm diff}$ is given in years. Our method of computing the
diffusion timescale includes the assumption that viscous torques provide
some angular momentum transport, but stellar radiation dominates the
disk energy budget. The effect of diffusion is most profound in the
inner disk where $t_{\rm diff}$ is smallest. The slightly reduced
surface density in the inner disk provides some additional gravitational
stability to the disk as a whole.

Figure \ref{diskfig} (bottom panel) shows the solid surface density
profiles of the maximum-mass and medium-mass nebulae for each star.
Each disk has an outer boundary of 100~AU. Gas/solid ratios are taken
from the disk molecular inventory at solar composition of
Dodson-Robinson et al.\ (2009). The maximum-mass disks hold between
$1/4$ and $1/3$ of the system mass and may undergo global gravitational
instabilities that aid planet/planetesimal formation (e.g.  Adams et
al.\ 1989; Rice et al.\ 2004; Lodato \& Rice 2005), which we examine in
\S \ref{gi}. Such massive disks would be either in the middle or just at
the end of their epoch of accretion from the protostellar clump,
consistent with our assumed stellar age of $10^5$ years. Although we do
not examine accretion-triggered instability, it is important to note
that the accretion rate onto the disk can be a determining factor for
fragmentation at large radii (Vorobyov \& Basu 2006; Boley 2009; Clarke
2009; Rafikov 2009). The medium-mass nebulae hold $< 10\%$ of the total
system mass, consistent with disk masses derived from sub-mm
observations (Andrews \& Williams 2007).

Having created our disk models, we now test the suitability of each for
forming planets.

\section{Experiment 1: Core Accretion}
\label{coreaccretion}

Since the astronomical community has reached a near-consensus that core
accretion is the dominant mode of planet formation---explaining such
observables as the planet-metallicity correlation (Gonzalez 1998;
Fischer \& Valenti 2005), the planet-silicon correlation (Robinson et
al.\ 2006), and the paucity of gas giants orbiting M dwarfs (e.g.
Johnson et al.\ 2007), we begin our investigation of planets on wide
orbits with an assessment of the ability to form critical-mass cores at
large radii. The quantity of interest is the maximum possible
protoplanet core mass as a function of distance from the star. If the
core can reach $10 M_{\oplus}$, the canonical critical value for
destabilizing the surrounding nebula to accrete a massive atmosphere
(Mizuno 1980), while the gas disk is still present, it has a chance of
forming a gas giant. We adopt a disk lifetime of 5~Myr, the maximum star
cluster age with an appreciable disk fraction according to Currie et
al.\ (2009). Here we are being generous: most disks have far shorter
observed lifetimes, but we must account for the possibility that planets
on wide orbits are relics of long-lived disks.

We use the feeding zone approximation (Safronov 1969; Lissauer 1993), in
which a solitary core accretes planetesimals at the rate
\begin{equation}
\dot{M} = \pi R^2 \Sigma \Omega \left ( 1 + \frac{2 G M}{R \langle v
\rangle^2} \right ) ,
\label{growthrate}
\end{equation}
where $M$ is the core mass, $R$ is the core radius, $\Sigma$ is the
surface density of planetesimals, $\Omega$ is the Keplerian frequency
and $\langle v \rangle$ is the RMS planetesimal velocity.  Assuming
solitary cores tends to increase the maximum planet core mass, since
oligarchic growth rates are much slower than those of a single embryo in
a swarm of planetesimals.  The feeding zone extends 4 Hill radii on
either side of the protoplanet (Kary \& Lissauer 1994), where the Hill
radius $R_H$ demarcates the protoplanet's Roche lobe in the spherical
approximation:
\begin{equation}
R_H = a \left ( \frac{M}{3 M_*} \right )^{1/3}.
\label{rhill}
\end{equation}
In Equation \ref{rhill}, $a$ is the planet's semimajor axis and $M_*$ is
the star mass. We begin the calculation with a $0.1 M_{\oplus}$ seed
core and use a fourth-order Runge-Kutta method to integrate Equation
\ref{growthrate} for 5~Myr, our adopted maximum disk lifetime. We
consider only the solid core growth and do not treat gas accretion.

To calculate the gravitational focusing factor (the ratio of the core
escape velocity to the mean planetesimal velocity), we apply the simple
assumption that $\langle v \rangle = R_H \Omega$, so that planetesimals
are at the Hill velocity. Accretion is therefore on the threshold
between dispersion-dominated, where only two-body forces (protoplanet
and planetesimal) are important, and shear-dominated, where the entire
star-planet-planetesimal system must be treated (Greenzweig \& Lissauer
1992).  Since three-body accretion rates tend to be lower than two-body
rates, our core accretion model may provide optimistically high
estimates of the core accretion efficiency. However, as we shall see,
core accretion cannot form gas giants at large radii even under the
extremely favorable conditions we have constructed here.

\begin{figure}
\epsscale{0.8}
\plotone{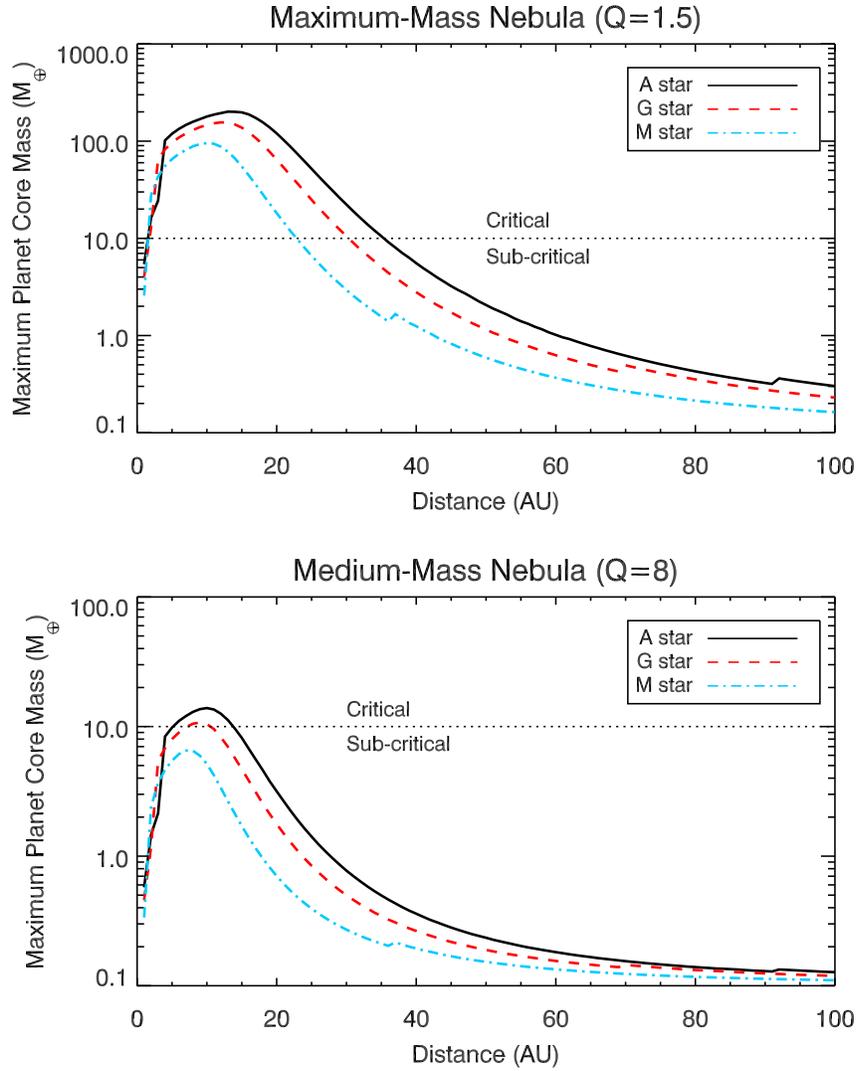}
\caption{Even our model A star hosting its maximum-mass nebula cannot
form gas giant planets by core accretion at distances larger than 35 AU.
{\bf Top:} Maximum planet core mass as a function of distance in the
maximum-mass nebulae. Assuming a critical core mass of $10 M_{\oplus}$,
the giant planet-forming region is restricted to $a < 35$~AU in the A
star disk, $a < 31$~AU in the G star disk, and $a < 23$~AU in the M star
disk. {\bf Bottom:} In the more typical medium-mass nebulae, giant
planet formation is restricted to a narrow region centered at $\sim
10$~AU in the model A and G star disks. The M star disk is not capable
of forming any gas giants by core accretion.}
\label{coregrowth}
\end{figure}

Figure \ref{coregrowth} shows the maximum planet core mass as a function
of distance from the star. In the medium-mass A and G star nebulae,
giant planet formation is comfortably confined to a narrow belt near
10~AU: there is no chance of forming a super-Jovian planet beyond 20~AU.
The maximum core mass barely reaches the critical $10 M_{\oplus}$ in the
G star disk, confirming that giant planet formation by core accretion is
a threshold phenomenon. A slight downward adjustment in disk mass,
metallicity or lifetime would be enough to prevent any giant planets
from forming---and indeed, the estimated planet occurrence rate between
0 and 5~AU is only 10\% for solar-mass stars (Cumming et al.\ 2008). The
medium-mass M star disk does not form any giant planets at all,
consistent with the observed paucity of gas giants orbiting M dwarfs
(Johnson et al.\ 2007).

We note that critical core mass is not constant throughout the disk, but
instead depends on the isolation mass of planetesimals in the feeding
zone. In the outer disk, where the isolation mass can reach $100
M_{\oplus}$ or more, the solid accretion rate remains high throughout
planet formation and the kinetic energy deposited by planetesimals
inhibits the protoplanet atmosphere's contraction. By using $10
M_{\oplus}$ as the critical core mass throughout our experiments, we are
overestimating the chances of giant planet formation in the outer disk:
simulations of planet formation in the trans-Saturnian solar nebula show
that critical mass can exceed $40 M_{\oplus}$ (Dodson-Robinson et al.\
2008; Dodson-Robinson \& Bodenheimer 2009).


Now, being extremely generous not only with the feeding zone setup and
disk lifetime, we investigate the core formation efficiency in the
maximum-mass nebulae. Since core accretion requires a stable, quiescent
disk, we cannot possibly load the disk with more planetesimal mass than
contained in the $Q = 1.5$ nebulae (except through second-order effects
such as high metallicity). If giant planet cores cannot reach critical
mass in our maximum-mass nebulae, they have no chance to do so in any
physically realistic disk. Figure \ref{coregrowth} shows that the
maximum radius where core accretion is effective is 35~AU in the A star
disk, 31~AU in the G star disk, and 23~AU in the M star disk. Core
accretion might be responsible for HR~8799~d, on a $\sim 23$~AU orbit
(Marois et al.\ 2008; Fabrycky \& Murray-Clay 2008), but the other
direct imaging discoveries HR~8799~b and c and Fomalhaut~b are well
outside the core accretion planet-forming zone.

Is it possible that a super-Jupiter could form on a $\sim 35$~AU orbit
and migrate {\it outward} to 50~AU or greater? Perhaps: there are two
robust outward migration mechanisms, planetesimal scattering and Type
III migration. In addition, Morbidelli \& Crida (2007) propose that two
giant planets in a 2:3 mean motion resonance that open overlapping gaps
in the disk may halt or reverse the normally inward Type II migration
process.

Hahn \& Malhotra (1999), followed by Gomes et al.\ (2004) and
Tsiganis et al.\ (2005), found that a $30-50 M_{\oplus}$ planetesimal
disk could expand Neptune's orbit from 23~AU to 30~AU. However, a planet
must interact with at least its own mass in planetesimals to change its
semimajor axis appreciably. Even our maximum-mass A star nebula contains
$< 2 M_{\rm Jup}$ of planetesimals between 35 and 100 AU. Since
HR~8799~b and c are each at least $7 M_{\rm Jup}$, we can rule out core
accretion followed by planetesimal-driven migration.

Type III migration, which is driven by torques at the planet's
corotation resonance and may proceed inward or outward, is the least
understood type of large-scale planet motion. However, Pepli\'{n}ski et
al.\ (2008) show that outward Type III migration (a) requires a sharp
inner disk edge located near the planet and (b) always reverses
direction and becomes inward migration after the planet's semimajor axis
has approximately doubled. Based on their work, we do not consider Type
III migration from the inner disk a viable explanation for the origins
of HR~8799~b and Fomalhaut~b.

Our first numerical experiment demonstrates that even under the most
favorable conditions possible---massive, long-lived disk; low-velocity
planetesimals; a seed core of $0.1 M_{\oplus}$ to start the
accretion---{\it gas giants cannot form by} in situ {\it core accretion
beyond 35~AU}.  For lower-mass G and M stars, the core accretion
scenario is even more unfavorable. The work of Pepli\'{n}ski et al.\
(2008) indicates that outward migration from the inner disk is also
implausible. The next two experiments investigate other ways of placing
super-Jovian planets on extremely wide orbits.

\section{Experiment 2: Scattering}
\label{scattering}

Next we consider the possibility that planetary systems such as HR 8799
may in fact have formed via core accretion---but as a system that was
initially more compact than currently observed and containing at least
one additional massive planet. (Here we use the term ``planet'' despite
the significant uncertainties in mass and formation mechanism).  The
unseen, massive planet scatters the discovered planets, forming in the
inner nebula according to the canonical core accretion theory, out to
their current wide orbits. Since HR~8799 is the only multi-planet system
that includes planets on ultra-wide orbits, we use it as a test case for
this experiment.

Our basic approach is to perform many N-body integrations of systems
with a 1.5$M_{\odot}$ A star, to match the mass of HR~8799, and four
planets. Our goal is to determine if planet scattering can lead to
systems qualitatively similar to the HR 8799 system. Three of the
planets correspond to the discovered objects, with masses of 7$M_{\rm
Jup}$, 7$M_{\rm Jup}$, and 5$M_{\rm Jup}$. These masses are
intentionally chosen to be near the lower limit of current mass
estimates (Marois et al.\ 2008), as previous studies have shown that
systems containing lower planet masses are less likely to be rapidly
disrupted by strong planet-planet scattering (Fabrycky \& Murray-Clay
2008; Go\'{z}dziewski \& Migaszewski 2009). The putative fourth
companion would need to have a mass equal to or larger than the other
planets so that it can effectively scatter the other planets outward
(Ford \& Rasio 2008).  Since the total mass in planets is already quite
large compared to other known planetary systems (Butler et al.\ 2006),
we set the mass of the putative fourth body to be $7M_{\rm Jup}$. We
integrate each initial configuration of planets for 200~Myr, using the
Mercury hybrid symplectic integrator (Chambers 1999).

The initial semi-major axis for the unseen inner planet is drawn from a
uniform distribution between 3 and 7~AU. The lower limit is chosen in
order to place the innermost planet as close to the star as it could
plausibly have formed via core accretion. As shown in Figure
\ref{coregrowth} (black curve), for A stars there is a rapid rise in the
maximum core mass near 3 AU in both the medium- and maximum-mass
nebulae. The upper limit for the initial semi-major axis of the inner
planet is based on the need for the outer planets to be close enough to
the central star for core accretion to remain a viable formation
mechanism even when the planets are placed several mutual Hill radii
apart (see below).

We assign initial semi-major axes for the three outer planets such that
the difference in the semimajor axis of each each pair of neighboring
planets is $a_{i+1}-a_i = K \times r_{H,i,i+1}$, where $K$ is a constant
of order unity and $r_{H,i,i+1}$ is the mutual Hill radius of the $i$th
and $i+1$th planet (Chatterjee et al.\ 2008):
\begin{equation}  
r_{H,i,i+1} = \left(\frac{M_i+M_{i+1}}{3M_*}\right)^{1/3} \frac{a_i+a_{i+1}}{2}
\end{equation} 
The choice of $K$ affects the spacing of the planets and thus the
timescale until the first close encounter. Based on similar scattering
experiments (but with three less massive planets), we expect that a
small choice for $K$ (e.g., $\sim~2-3.5$) is likely to result in a very
rapid first scattering event, which would cast doubt on the plausibility
on the initial conditions. A somewhat larger choice of $K\sim~4$ allows
the system to evolve for hundreds or thousands of orbital periods before
the first close encounter. Thus, the scattering is triggered by
dynamical instability in a chaotic system rather than the choice of
initial conditions (Chatterjee et al.\ 2008). In this case, we cannot
incrase $K$ much further, as the initial semi-major axis of the
outermost planet becomes prohibitively large in the context of formation
via core accretion (Figure 2).

Each planet is assigned an initial eccentricity of less than 0.05, an
initial inclination of less than 0.025 radians, and random angles for
the initial pericenter direction, initial longitude of ascending node
and initial mean anomaly.  We perform two sets of $\simeq$300
simulations, one with $K=4.0$ and another with $K=4.5$. The sequence of
the three observed planets is chosen randomly. Each planet is assigned
an initial eccentricity of less than 0.05, an initial inclination of
less than 0.025 radians, and random angles for the initial pericenter
direction, initial longitude of ascending node and initial mean anomaly.

The median initial semi-major axis of the outer-most planet is $33.8 \pm
5.0$~AU for $K=4.0$ and $40.2 \pm 7.0$~AU for $K=4.5$.  Thus, three of
the simulated planets typically start at smaller semi-major axes than
the observed separations for all three observed companions. One
simulated planet typically has an initial semi-major axis between HR
8799 c and d in the $K=4.0$ simulations, or comparable to the present
separation of planet c in the $K=4.5$ simulations. As shown in Figure
\ref{coregrowth} (black curve), the maximum core mass cannot reach
critical core mass beyond separations of $\sim 35$~AU. If taken
literally, this would suggest that only our simulations with an inner
planet initially near 3~AU could result in four giant planets with each
separated from its neighbors by at least 4 mutual Hill radii.
Nevertheless, we consider initial conditions that include systems with
somewhat wider initial spacings, given the large current separations of
planets HR8799 b, c and d. While the initial conditions with $K=4.0$ are
more consistent with the modeling in \S \ref{coreaccretion}, most of
these simulations result in rapid instabilities (within $\sim$1Myr or
$\sim$5,000 orbital periods of the outer planet). This motivated the
additional simulations with $K=4.5$ which had their first close
encounter on a longer timescale of $\sim$24Myr, near the minimum age of
the host star (30~Myr; Marois et al.\ 2008).

For the simulations with $K=4.5$, $\simeq 32\%$ of the systems retained
exactly three planets at the end of the integrations. With $K=4.0$, the
retention rate was far worse: $\le~1\%$ of the systems retained exactly
three planets. In both cases, the systems that retained three planets
did not have architectures similar to HR 8799, since one of the planets
had migrated inwards and thus only two remained at large separations.
Some of these systems did not suffer an orbital instability (so no
planets at separations comparable to HR 8799b), while others had planets
with final semi-major axes quite different from those of HR 8799 (e.g.,
outermost planet with a semi-major axis much greater than the other
planets). Thus, our simulations suggest that planet scattering alone
cannot produce a stable system that looks like HR 8799---although
Fomalhaut b, with a solitary planet on a wide orbit, might be possible.

The above result leads us to consider the possibility that HR 8799, and
future discoveries of multiple, massive planets on wide orbits, might be
dynamically {\em unstable}. Given the relatively young age of the host
star ($< 160$~Myr; Marois et al.\ 2008), we might be observing the
planetary system in a transient phase while planet-planet interactions
are still acting to resolve a recent dynamical instability
(Go\'{z}dziewsi \& Migaszewski 2009). Thus, we analyzed snapshots of
each system taken every 200,000 years up to a maximum duration of 160
Myr. We include snapshots at times as small as 200,000 year after the
beginning of our n-body integrations.

We identify those snapshots with three outer planets that have
semi-major axes between 20 and 100 AU, qualitatively similar to HR 8799
b, c and d, plus a fourth planet that is closer to the star and would
have eluded detection.  For the simulations with $K=4.0$ (4.5), we find
$\sim0.01\%-0.07\%$ ($\sim0.01\%-0.02\%$) of snapshots meet this
criteria depending on whether we look at the entire 160~Myr or just the
first 30~Myr of our simulations. These simulations suggest that if the
HR 8799 system is the result of core accretion at smaller separations
followed by planet-planet scattering, then we would be observing the
system during a very brief transient phase of its evolution. {\it Thus,
planet scattering is not a viable formation model if systems such as HR
8799 are found to be common.} However, at least for the time being, HR
8799 is the only multi-planet system discovered by direct imaging and
likely subject to strong selection effects.

In summary, we cannot conclusively rule out the model of core accretion
followed by planet scattering for young systems that might still be
dynamically evolving.  Our results are consistent with the findings of
Go\'{z}dziewski \& Migaszewski (2009), Scharf \& Menou (2009) and Veras
et al.\ (2009), who find dynamical relaxation timescales of only 10~Myr
for tightly packed systems. Indeed, Scharf \& Menou report a planet
retention rate of $20\%$ after 10~Myr. Furthermore, Raymond et al.\
(2009) show that while planetesimals may help retain planets that would
otherwise have been ejected, the eccentricity distribution of massive
scattered planets remains wide---contrary to the case of HR~8799 where
stability requires nearly circular orbits (Lafreni\`{e}re et al.\ 2009).
Even Fomalhaut b, with a minimum eccentricity of 0.13 (Kalas et al.\
2008), cannot have a periapse-apoapse swing of more than 15~AU without
disrupting its neighboring debris disk. {\it We recommend that observers
examine the planet detection rate as a function of stellar age,
controlling for the planets' dimming with time.} If planet detection
rate is found to be independent of stellar age, it would confirm our
finding that planet scattering is not the dominant mode of producing
detectable planets on wide orbits.

Finally, there is the possibility---not explored here---that large
planet masses may allow for significant planet-planet interactions to
begin occurring while there are still significant planet-disk
interactions (e.g., migration, eccentricity damping, and accretion) that
are not included in our model. Unfortunately, detailed models of
planet-disk interactions over $\sim$100~Myr would be extremely
computationally demanding and are thus left for a future study.

\section{Experiment 3: Gravitational Instability}
\label{gi}

Our planet scattering experiment was not successful at making a stable
system with massive planets on wide orbits, and we showed in \S
\ref{coreaccretion} that core accretion plus migration have severe
difficulties producing massive gas giants on wide orbits. In this
section we investigate the one remaining possibility, gravitational
instability in the disk. Following (1) the breakup of a disk into
fragments; (2) the orbital evolution of each fragment; (3) the
fragments' late-stage accretion of both planetesimals and gas; and (4)
the contraction of a fragment to approximately Jupiter's radius is
beyond the capability of any single numerical model.  As a first step we
simply verify that our model disks can become gravitationally unstable.

\subsection{Mathematical Method}

We search for exponentially growing spiral modes using the linear
stability analysis method of Adams et al.\ (1989), further developed by
Laughlin \& Rozyczka (1996). The method is described extensively in the
literature, so we provide only an overview here.

The disk surface density is decomposed into an unperturbed component
plus an exponentially growing mode in spiral form:
\begin{equation}
\Sigma(r, \theta, t) = \Sigma_0(r) + \Sigma_1(r) \exp [i (\omega t - m
\theta )] .
\label{mode}
\end{equation}
In Equation \ref{mode}, $\Sigma_0$ is the unperturped surface density,
$\Sigma_1$ is the perturbed component, $m$ is the number of spiral arms,
and ($r$, $\theta$, $t$) are the radial, azimuthal and time coordinates.
The complex eigenfrequency $\omega$ is given by
\begin{equation}
\omega = m \Omega_p - i \gamma ,
\label{eigen}
\end{equation}
where $\Omega_p$ is the fixed pattern speed and $\gamma$ is the growth
rate of the spiral wave. The disk radial velocity, azimuthal velocity
and potential $\psi$ due to self-gravity can be modally decomposed in an
analogous manner to the surface density. Unperturbed quantities
$\Sigma_0$ and $c_s^2$ (sound speed) are specified by the disk models
described in \S \ref{diskmodel}. Here we assume that the disk is
vertically isothermal, consistent with the Chiang \& Goldreich (1997)
passive disk model adopted in \S \ref{diskmodel}.


We begin our search for growing modes by calculating the rotation
curve---including self-gravity and pressure support---and corresponding
epicyclic frequency:
\begin{equation}
\begin{array}{c}
\bigskip
\displaystyle
r \Omega^2(r) = \frac{G M_*}{r^2} + \frac{c_s^2}{\Sigma_0} \frac{d
\Sigma_0}{dr} + \frac{d}{dr} \left (\psi_0 \right) , \\
\bigskip
\displaystyle
\kappa^2(r) = \frac{1}{r^3} \frac{d}{dr} \left [ \left ( r^2 \Omega
\right )^2 \right ]
\end{array}
\label{rotcurve}
\end{equation}
The disk self-gravity term, the third term on the right-hand side of
Eq.\ \ref{rotcurve}, is numerically evaluated as
\begin{equation}
r \Omega^2_{disk} =  -\frac{d}{dr} \int_0^{2 \pi} d \theta \int_{R_{\rm
in}}^{R_{\rm out}} \frac{G \Sigma_0(r') r' dr'}{\left ( r^2 + r'^2 - 2 r
r' cos \theta + \eta^2 \right )^{1/2}} ,
\label{omdisk}
\end{equation}
where the softening parameter $\eta = 0.1$ allows the azimuthal
integral, which diverges at the disk edges, to be evaluated.

The next step is to calculate the potential due to self-gravity of the
growing spiral perturbation:
\begin{equation}
\psi_1 = -G \int_{R_{\rm in}}^{R_{\rm out}} \Sigma_1(r') r' dr'
\int_0^{2 \pi} \frac{d \theta}{\sqrt{r^2 + r'^2 - 2 r r' \cos
\theta}} .
\label{poisson}
\end{equation}
We define the perturbed component of the disk enthalpy, which works
against the growth of the spiral mode, as
\begin{equation}
h_1 = c_s^2 \frac{\Sigma_1}{\Sigma_0} .
\label{enthalpy}
\end{equation}
The linear stability analysis method is built on the approximation that
density waves propagate only radially and azimuthally. Such a
formulation requires that we treat the disk as a two-dimensional
surface, so we cannot use volume density in our calculations.
Therefore, although we have introduced enthalpy into the calculation to
treat adiabatic density perturbations, we use the isothermal sound
speed, which does not depend on density, in Equations \ref{rotcurve} and
\ref{enthalpy}.

Finally, we combine the linearized equations of motion (continuity,
radial force and azimuthal force; written out in full in Adams et al.\
1989) into a single homogeneous integro-differential equation in terms
of $h_1$ and $\psi_1$:

\begin{equation}
\frac{d^2}{dr^2} \left( h_1 + \psi_1 \right ) + A \frac{d}{dr}
\left (h_1 + \psi_1 \right ) + B \left (h_1 + \psi_1 \right ) + C h_1 =
0,
\label{governing}
\end{equation}
with the coefficients $A$, $B$, $C$ and $\nu$ defined as
\begin{equation}
\begin{array}{l}
\displaystyle
\bigskip
A = \frac{d}{dr} \log \left [ \frac{\Sigma_0 r}{\kappa^2 (1 -
\nu^2)} \right ], \\
\bigskip
\displaystyle
B = -\frac{m^2}{r^2} - \frac{4m}{r^2} \frac{\Omega}{\kappa}
\frac{r}{(1 - \nu^2)} \frac{d \nu}{dr} + \frac{2m}{r\nu}
\frac{\Omega}{\kappa} \frac{d}{dr} \log \left ( \frac{\kappa^2}{\Omega
\Sigma_0} \right ), \\
\bigskip
\displaystyle
C = -\frac{\kappa^2 (1 - \nu^2)}{c_s^2}, \\
\bigskip
\displaystyle
\nu \equiv \frac{\omega - m \Omega}{\kappa}. 
\end{array}
\label{coeffs}
\end{equation}
For the full derivation of the governing equation, see Adams et al.\
(1989).

The goal of linear stability analysis is to find combinations of $m$,
$\Omega_p$ and $\gamma$ that satisfy Equation \ref{governing}. Such
combinations represent exponentially growing spiral modes. Since we do
not {\it a priori} know the perturbed surface density $\Sigma_1$
associated with any mode, we discretize the disk and write the governing
equation in matrix form:
\begin{equation}
{\mathcal M}_{jk} \Sigma_{1_k} = 0 .
\label{mateq}
\end{equation}
In Equation \ref{mateq}, $\Sigma_{1_k}$ is the perturbed surface density
at position $k$ of a discrete, linear grid.  The matrix ${\mathcal
M}_{jk}$ acts on the perturbed surface density to evaluate the left-hand
side of Equation \ref{governing}. Since Equation \ref{governing} is
homogeneous, it has a solution if the complex eigenfrequency $\omega$
(Equation \ref{eigen}) is an eigenvalue of
${\mathcal M}_{jk}$, such that
\begin{equation}
{\rm det} | {\mathcal M}_{jk} | = 0.
\label{determ}
\end{equation}

To construct ${\mathcal M}_{jk}$, we identify the rows with variable $r$
in Equation \ref{poisson} and the columns with variable $r'$. ${\mathcal
M}_{jk}$ must encode a method of integrating Poisson's equation (Eq.\
\ref{poisson}), which is singular on the diagonal, to calculate
$\psi_1$.  For the diagonal elements of ${\mathcal M}_{jk}$, we use
open-interval Romberg integration (Press et al.\ 1992) to evaluate the
azimuthal integral. We incorporate coefficients of the extended
trapezoidal rule to perform the integration over $r'$. The top and
bottom rows of matrix ${\mathcal M}_{jk}$ encode the boundary conditions
describing the flow of gas at the inner and outer disk edges. We follow
Laughlin \& Korchagin (1996) in requiring the radial component of the
velocity to vanish at both boundaries. The effect of these boundary
conditions is to forbid the disk from expanding inward or outward, which
results in lossless reflection of density waves at the disk boundaries.
The rigid boundary approximation and the value of the softening
parameter $\eta$ in Eq.\ \ref{omdisk} combine to artificially affect
stability at the disk edges (Adams et al.\ 1989).


Finally, we need a way of honing in on eigenvalues that provide
solutions to Equation \ref{governing}. We restrict our analysis to
two-armed spirals and do not consider higher-order modes. Viable pattern
speeds have their corotation resonances inside the disk, so we check
values of $\Omega_p$ that lie along the rotation curve (Equation
\ref{rotcurve}). We desire modes that grow within a few to a few hundred
orbital periods, so they can substantially modify the structure of the
disk before accretion or stellar heating stabilizes it: the growth rate
is therefore constrained by $0.01 \Omega_p \la \gamma \la \Omega_p$. We
use the Newton-Raphson method to identify the valid eigenmodes beginning
with gridded guesses for $\Omega_p$ and $\gamma$. Once we have found a
solution to Equation \ref{determ}, we set the first component of the
perturbed surface density vector to unity and use least-squares analysis
to solve the overconstrained system resulting from Equation \ref{mateq}.

\subsection{Results}

\begin{figure}
\centering
\begin{tabular}{ccc}
\includegraphics[scale=0.4]{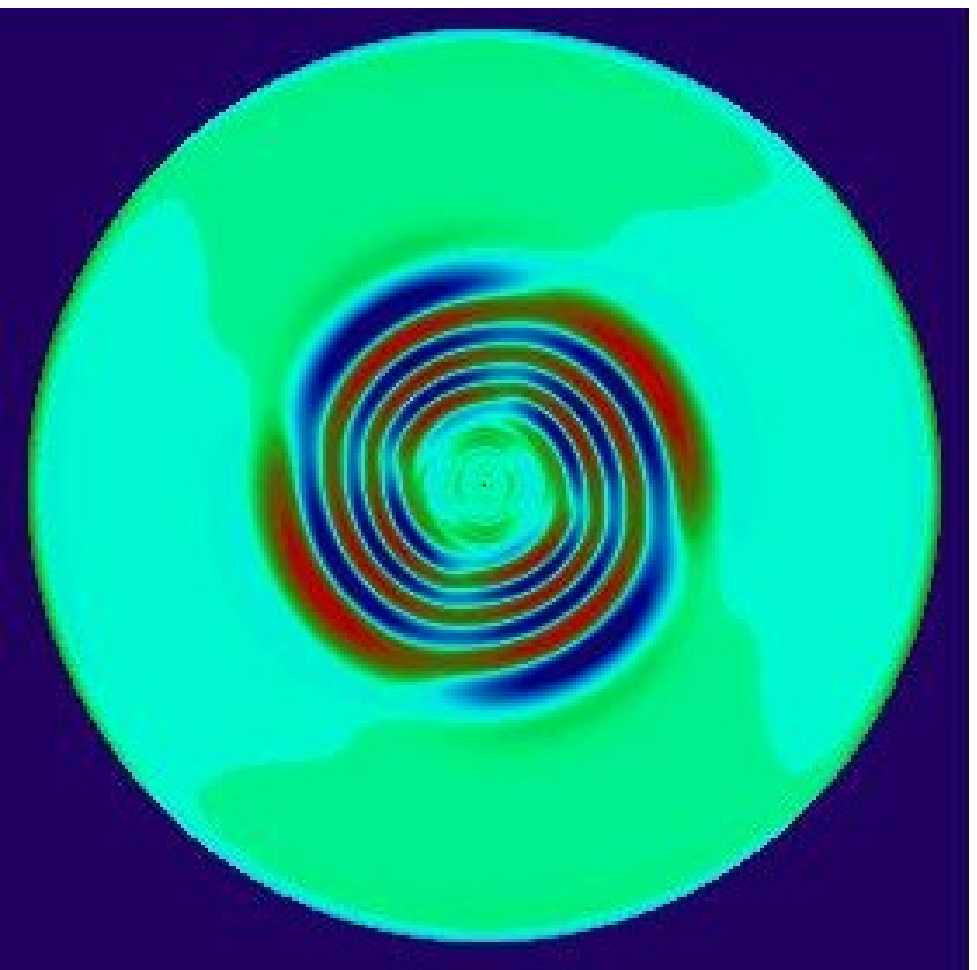} &
\includegraphics[scale=0.4]{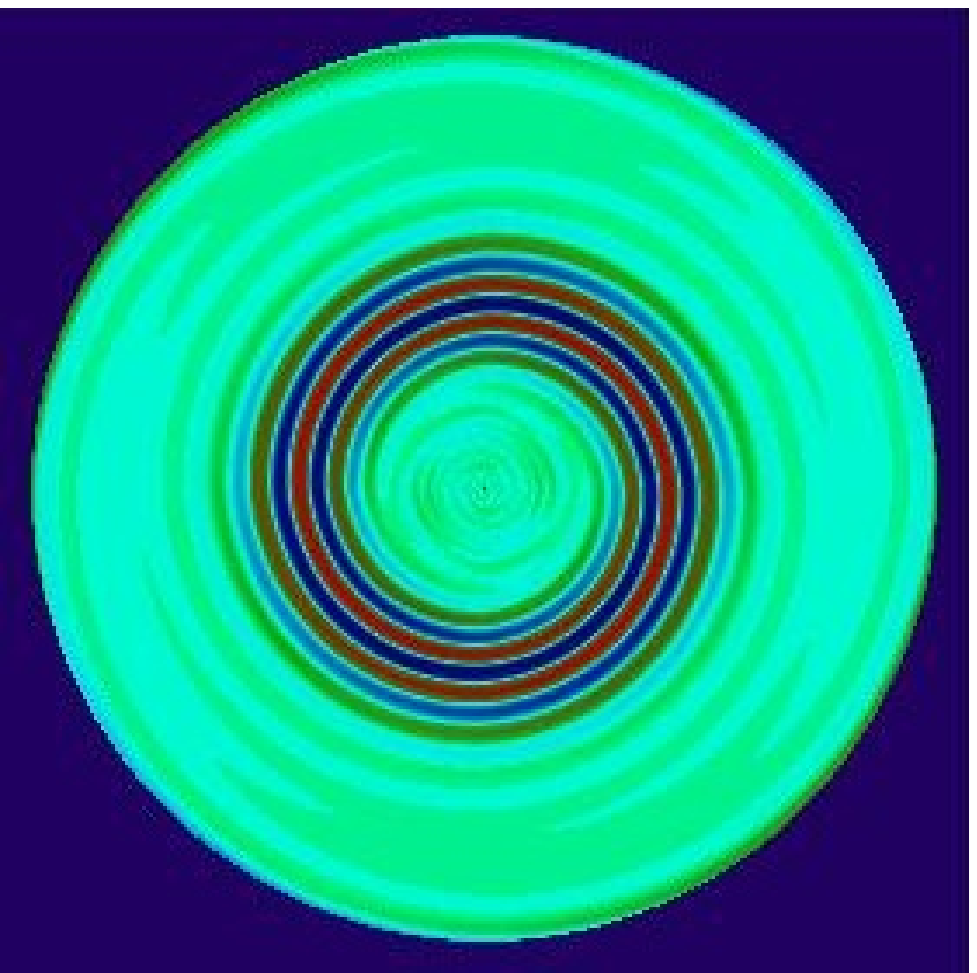} &
\includegraphics[scale=0.4]{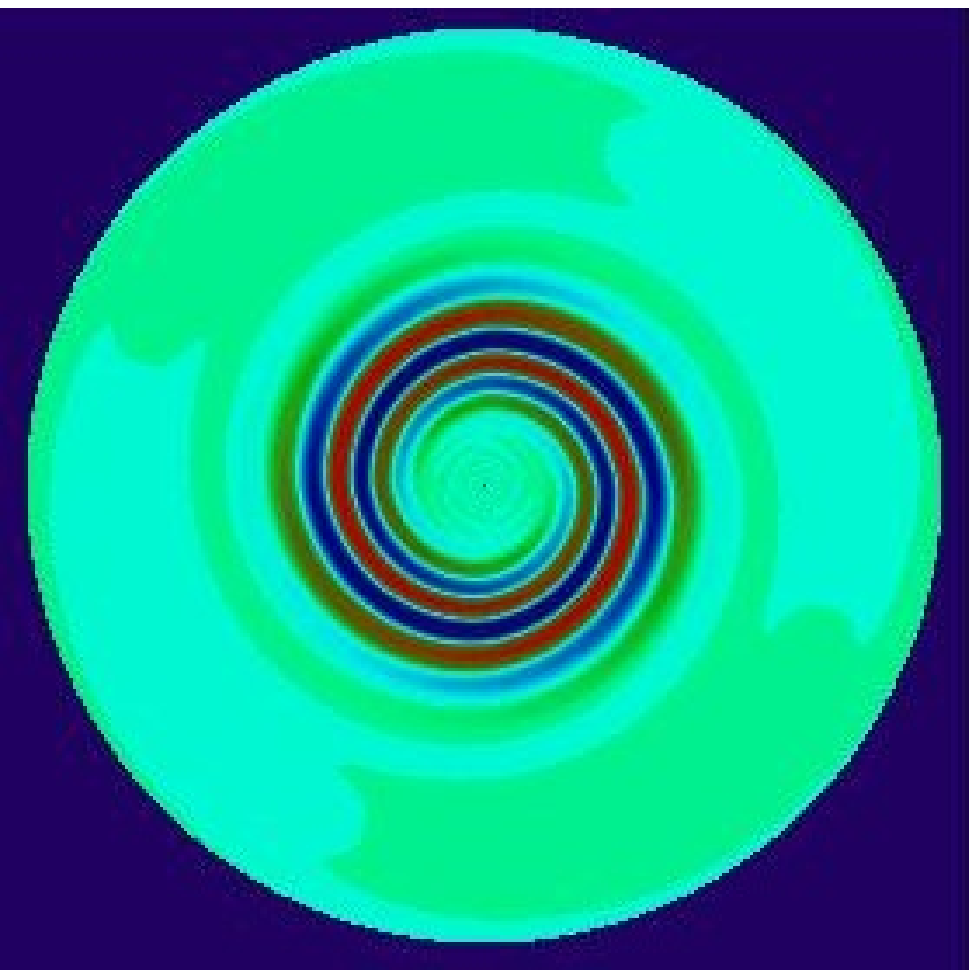} \\
\includegraphics[scale=0.4]{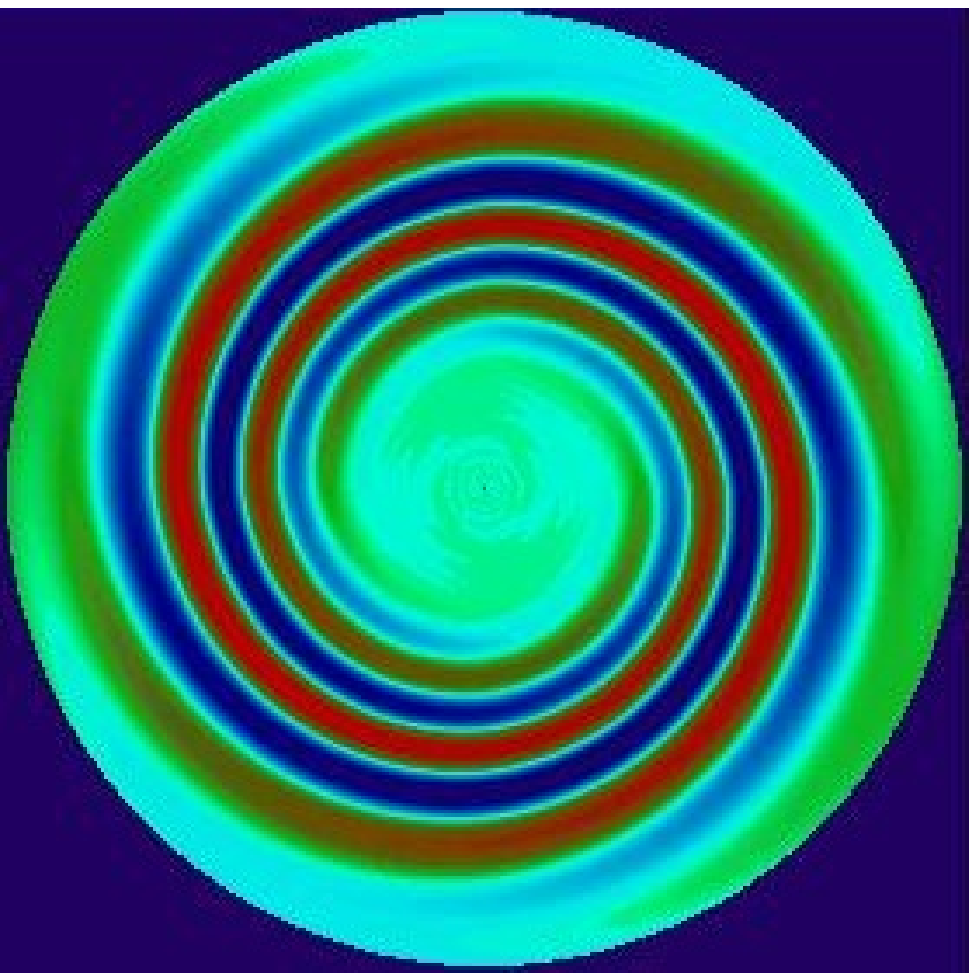} &
\includegraphics[scale=0.4]{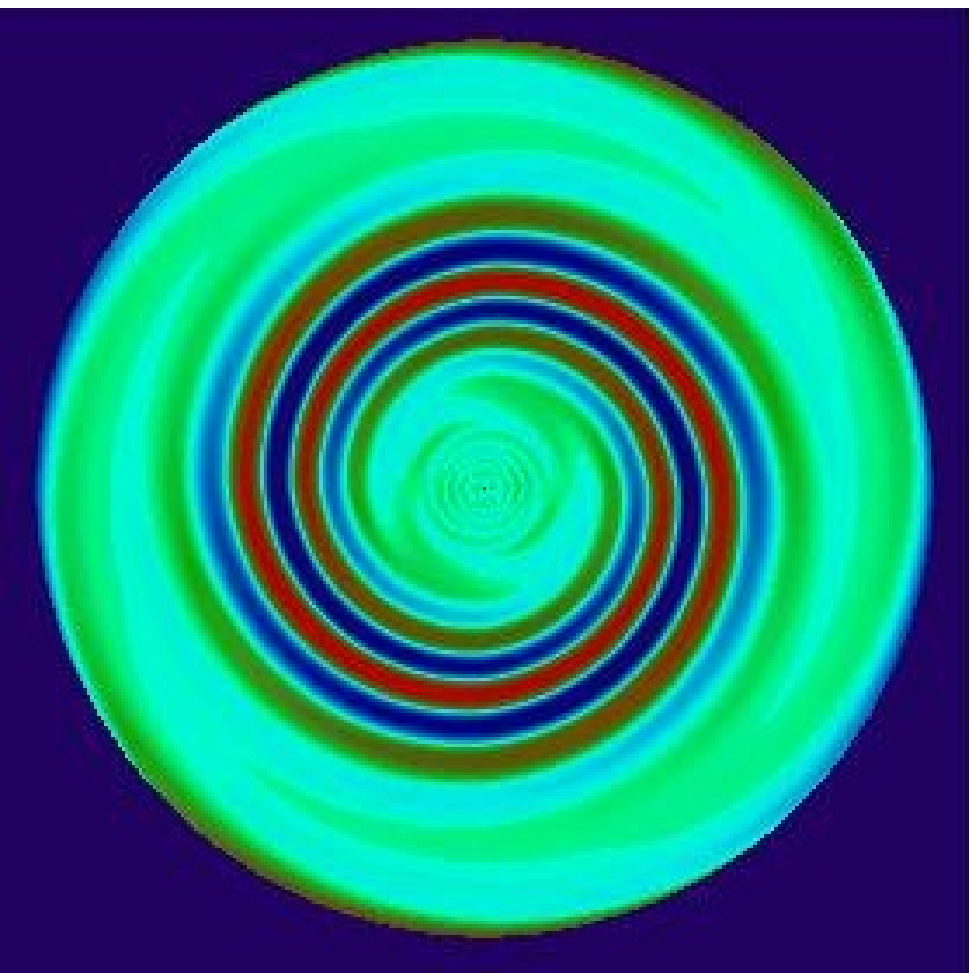} &
\includegraphics[scale=0.4]{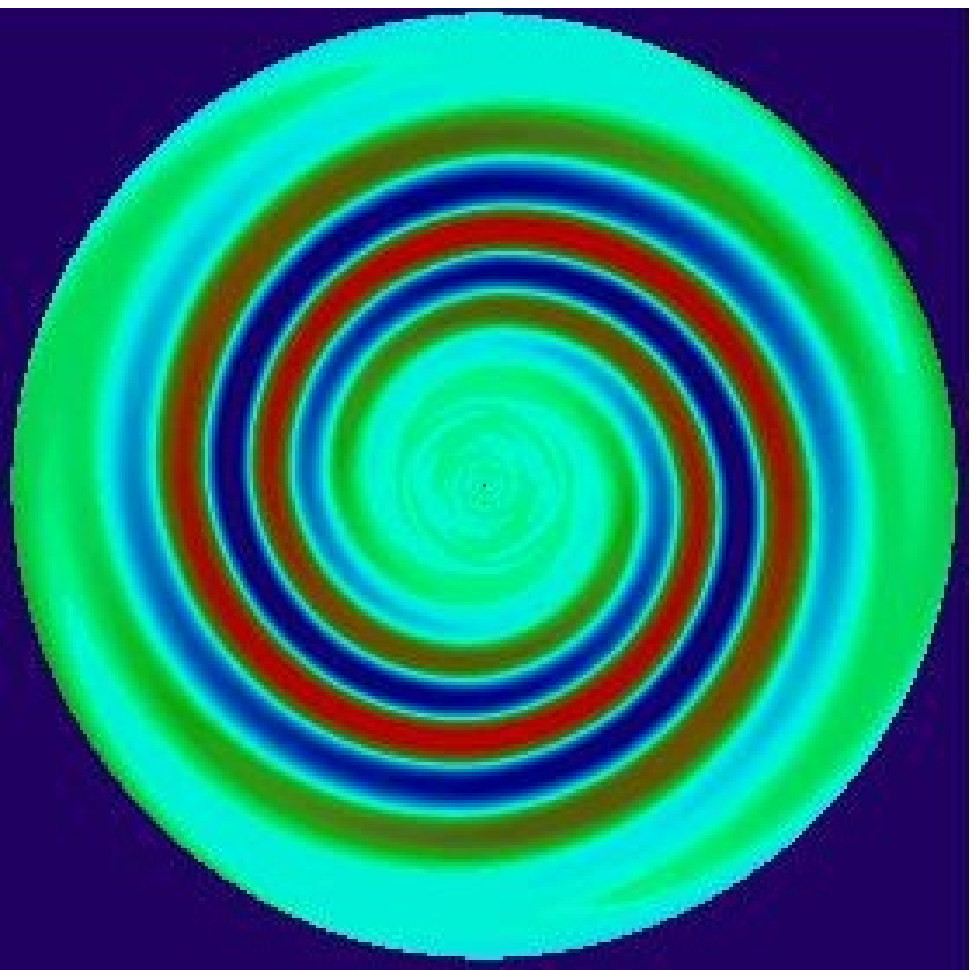} \\
\end{tabular}
\caption{The maximum-mass nebulae are unstable to exponentially growing
$m = 2$ spiral waves even though they are locally stable to axisymmetric
perturbations. The left column shows growing eigenmodes in the A star
maximum-mass nebula, the middle column shows growing modes in the G star
nebula, and the right column shows spiral structure in the M star
nebula. The color scale proceeds smoothly from red in the overdense
regions, to green in the unperturbed regions, to blue in the underdense
areas. Maximum surface density perturbations straddle the corotation
resonances, located between 15 and 53~AU from the star for the
instabilities shown here. {\it Our numerical experiments demonstrate
that gravitational instability is the only viable planet formation
mechanism for planets on wide} ($a \ga 35$~AU) {\it orbits.}}
\label{spirals}
\end{figure}

Our linear stability analysis reveals that the maximum-mass nebulae do
indeed undergo global gravitationally instabilities. In all the
maximum-mass disks, regardless of host star mass, we found multiple $m =
2$ growing modes with corotation resonances inside the disk. We list the
fastest-growing modes uncovered by the Newton-Raphson algorithm---those
with growth rates greater than $10\%$ of the orbital speed at the outer
disk edge---in Table \ref{modes}. For each mode the maximum-amplitude
perturbations straddle the corotation resonance and are contained within
the inner and outer Lindblad resonances, which occur at $\Omega_p =
\Omega \mp \kappa/2$. Figure \ref{spirals} shows six of the modes listed
in Table \ref{modes}. Units on the color scale are arbitrary as the
modes grow exponentially. In our simple formulation, where the disk
characteristics are completely determined by the stellar radius and
temperature, the viability of spiral modes does not appear to depend on
the mass of the star: any system with a high disk/star mass ratio can
become unstable.

Since our maximum-mass disks are both thin and massive relative to their
host stars, containing between $1/4$ (A star) and $1/3$ (M star) of the
total system mass, the fact that they become unstable is not at all
surprising. Numerous previous investigations, including Adams et al.\
(1989), Laughlin et al.\ (1998), Lodato \& Rice (2005), Kratter et al.\
(2008), Cai et al.\ 2008, and Stamatellos \& Whitworth (2009), have
verified that cold disks with more than 10\% of the mass of their stars
are subject to long-wavelength, global gravitational instabilities.

Again unsurprisingly, we were not able to find growing modes in our
medium-mass nebulae within the limitations of a grid-based search for
modes using the Newton-Raphson scheme. If planets on wide orbits form by
gravitational instability, they can do so only in massive disks. They
must be relics of the early (Class 0--Class 1) stages of protostar
evolution, where massive disks are still present.

At long last one of our experiments has revealed that there is a chance
of forming the companions to HR~8799 and Fomalhaut, and placing them on
stable orbits, by a known mechanism. However, there is still a lot of
work to be done to confirm that these planets are products of
gravitational instability. Our experiments have merely verified that our
maximum-mass disks can become unstable. We do not yet know what happens
to the resulting spiral modes. We must rely on the work of previous
investigators, whose simulations have shown spiral arms in massive disks
can fragmenting into self-gravitating clumps of several Jupiter masses
(e.g.\ Boss 2000; Pickett et al.\ 2003; Mayer et al.\ 2004; Boley 2009).
Even more importantly, simulations by Mayer et al.\ (2004) and Boss
(2005) indicate that the protoplanetary clumps can survive without
migrating into the star.

We caution that the viability of disk instability as a planet formation
mechanism has not been fully established. The fragment survival issue
requires further investigation, as Mej\'{i}a et al.\ (2005) show that
clumps may be sheared apart within an orbit or less. Pickett \& Durisen
(2007) caution that clump longevity in numerical simulations strongly
depends on artificial viscosity, which is implemented differently in
smoothed particle hydrodynamics and grid-based calculations. Boley et
al.\ (2007) and Cai et al.\ (2009) improve upon the radiative transfer
treatment of Boss (2007), who found rapid disk cooling and
fragmentation, and recover a cooling rate too slow for {\it any}
fragmentation to occur over several outer disk rotations. The emerging
consensus seems to be that while fragmentation cannot occur in the inner
disk, $a \la 20$--40~AU (Rafikov 2005; Boley \& Durisen 2008;
Stamatellos \& Whitworth 2008; Cai et al.\ 2009; Forgan et al. 2009), it
does occur in extended disks outside of a few tens of AU (Boley 2009;
Stamatellos \& Whitworth 2009) provided that the disks are massive. The
analytical work of Clarke (2009) demonstrates that disks accreting from
the protostellar core inevitably fragment upon reaching a radial extent
of 70~AU.


At this point we have only circumstantial evidence that massive planets
on wide orbits, beyond 35~AU, form by gravitational instability. In this
work disk instability emerges as the only possible theory from a
``survival of the fittest'' experiment, but it is possible that some
fundamental piece of physics is missing from our understanding of planet
formation. Nevertheless, unless the wide-orbit planetary systems we have
observed are dynamically unstable---unlikely given their apparently
near-circular orbits---it seems like there is no other possibility. Core
accretion at large stellar separations unequivocally does not work.
{\it Our numerical experiments demonstrate that gravitational
instability is the only viable planet formation mechanism for planets on
wide} ($a \ga 35$~AU) {\it orbits.} We predict that HR~8799 b, c and d,
Fomalhaut b, and similar planets that may yet be discovered are the
result of gravitational instabilities in protostellar disks.

Boley (2009) argues for two modes of planet formation, with core
accretion dominating near the star and disk instability forming a
population of gas giants at large radii. To Boley's conclusion that
super-Jupiters on wide orbits should be a significant, or even dominant,
component of the gas giant population of low-metallicity stars---in
whose disks core accretion is inefficient---we add that the same is true
of {\it low-mass} stars. Figure \ref{coregrowth} demonstrates that gas
giant formation by core accretion is all but impossible for M dwarfs.
Core accretion has a chance only in M-star disks with extremely high
mass and/or metallicity, but high-mass disks are likely to fragment and
destroy the quiescent feeding zones necessary for steady growth by core
accretion.

Clarke (2009) argues that companions formed by disk fragmentation
typically have at least $10\%$ of the primary star mass. The
minimum-mass disk-formed companion to a $0.1 M_{\odot}$ star would
therefore be 10 Jupiter masses and would be classified as a planet
rather than a brown dwarf. {\it We predict that the occurrence ratio of
long-period to short-period gas giants should be highest for M dwarfs},
reflecting both the inefficiency of short-period planet formation by
core accretion and the planetary-scale minimum mass of wide-orbit
fragments formed by disk instability.

Having finally identified a plausible formation mechanism for massive
planets in wide oribts, we conclude our article with a review of our
findings and recommendations for observers on how to confirm the
gravitational instability hypothesis.

\section{Conclusions and Testable Predictions}
\label{conclusions}

We have investigated the likelihood of forming massive planets on wide
orbits by three methods: near-{\it in situ} core accretion, scattering
by an unseen companion, and gravitational instability. Highlighting the
theoretical difficulties presented by the companions to Fomalhaut and
HR~8799, the prospects for forming such planets are generally poor. We
can certainly rule out core accretion, as it is hard to imagine a way
around the fact that long dynamical times destroy the chances of
building critical-mass cores. So far, numerical simulations of Type III
migration do not indicate that there is any possibility of moving a
planet all the way from 35~AU---the largest possible stellar separation
at which gas giants can form by core accretion---to $> 60$~AU
(Pepli\'{n}ski et al.\ 2008).

Planet-planet scattering can at least move a planet into a wide orbit;
the difficulty is keeping it there. Our N-body simulations (\S
\ref{scattering}) demonstrate that once scattered, planets tend to leave
the system. Planetesimals---not included in our calculations---can
stabilize the orbits of scattered planets through dynamical friction.
However, a planet needs to interact with of order its own mass in
planetesimals to circularize its orbit. Even our maximum-mass A star
nebula, with total mass $0.51 M_{\odot}$ within 100~AU, does not contain
7--10~$M_{\rm Jup}$ of planetesimals beyond 35~AU. If HR~8799 is an
unstable system, planet-planet scattering may be responsible for its
architecture, but the shepherded debris disk surrounding Fomalhaut
(Kalas et al.\ 2005) makes its present dynamical instability unlikely.

The one planet formation mechanism that can succeed at creating massive
gas giants on wide, near-circular orbits is gravitational instability.
In \S \ref{gi} we demonstrated that the same maximum-mass disk models
which could not form planets on wide orbits by core accretion do undergo
global spiral instability. {\it We predict that massive gas giants on
wide orbits are the products of protostellar disk fragmentation.} This
prediction, however, comes with the caveat that the detailed physics of
the disk instability process must be further analyzed and the clump
shearing and disk cooling time problems resolved (Mej\'{i}a et al.\
[2005]; Cai et al.\ [2006]; Pickett \& Durisen [2007]; Rafikov [2009]).

A simple observational test would resolve the question of how massive
planets on wide orbits form. So far 11 planets or Y dwarfs orbiting
main-sequence stars have been discovered by direct imaging. Once the
number of direct imaging discoveries reaches $\sim 30$, we can begin to
look for simple statistical patterns in the properties of planet hosts.
We propose dividing the stars surveyed by direct imaging into young and
old age bins, with the dividing age of order $\sim 100$~Myr (the longest
relaxation time in our simulations), and calculating the wide-orbit
planet occurrence rate of young vs.\ old stars. {\it If massive planets
in wide orbits are equally common around young and old stars,
gravitational instability is almost certainly their formation
mechanism.} Any age trend would indicate that planets on wide orbits are
transient relics of scattering from the inner disk. In conducting such
an experiment, care must be taken to control for the fact that planets
cool and become less detectable over time.

Further evidence for dual modes of planet formation, with core accretion
operating near the star and gravitational instability dominant at large
distance, would come from examining the occurrence ratio of long-period
to short-period gas giants as a function of stellar mass. We predict
that this ratio should be highest for M dwarfs, for which core accretion
is inefficient, short-period planets are rare (Johnson et al.\ 2007;
Cumming 2008), and disk fragmentation appears to produce objects of
planetary mass (Clarke 2009). Note, however, that preferred fragment
mass in extended disks is still a subject of considerable debate (e.g.
Boley et al.\ 2009, Kratter et al.\ 2009). Core accretion on its own
would produce a planet population where the occurrence ratio of
long-period to short-period gas giants has the opposite trend,
increasing with stellar mass.

As further evidence for gravitational instability, we note that HR~8799
is a $\lambda$~Bootis star, a metal-poor Pop I star that probably
accreted a thin layer of low-metallicity gas after formation (Marois et
al.\ 2008). Even if the protoplanetary disk surrounding HR~8799 was not
originally massive, passing through a nearby cloud could have triggered
Bondi-Hoyle accretion onto the disk and destabilized it (Throop \& Bally
2008).

One further discriminant of planets formed by disk instability and core
accretion/scattering is near-infrared color. Fortney et al.\ (2008) find
that model planets with the $\sim 5 \times$ Solar metallicity of Jupiter
and Saturn---generally interpreted as a signature of core
accretion---are redder by 1.5 mag in H-K than solar-metallicity models.
Note, however, the simulations by Helled \& Schubert (2008) showing that
planets formed by disk instability may later be enriched by planetesimal
capture.

Finally, direct detection of planets is still a young enough field that
finding as many planets as possible, without regard to the properties of
their host stars, is a worthwhile goal. Gravitational instability
naturally produces planets on wide orbits (20--150~AU, Mayer et al.\
2004; Boley 2009; Clarke 2009; Rafikov 2009; Stamatellos \& Whitworth
2009), but the core accretion zone of proto-A stars extends to 35~AU and
can be directly imaged for nearby targets. Our work confirms the finding
of Kennedy \& Kenyon (2008) that A stars have wider planet formation
zones than lower-mass stars. Lovis \& Mayor (2007) also found the
highest average planet mass around A stars vs.\ G and M stars, which
aids planet detectability. Furthermore, A star disks experience the
highest infall rates during formation and are better candidates than
lower-mass G or M stars for triggered gravitational instability (Kratter
et al.\ 2009). We do not believe it is a coincidence that HR~8799 and
Fomalhaut are both A stars. {\it Where the goal is simply to image
planets, A stars make the best targets.} Note that here we are
distinguishing between the bulk planet detection likelihood---which so
far is highest for A stars in both radial velocity and direct imaging
surveys---and the {\it ratio} of long-period to short-period planet
occurrence.

Funding for S.D.R.'s work was provided by NASA through the Spitzer Space
Telescope Fellowship Program. E.B.F. and D.V. received support from the
National Science Foundation under the NSF grant listed below and the
University of Florida under the auspices of U.F.'s High-Performance
Computing Center. This project was the outgrowth of discussions at the
2009 Florida Astrophysics Winter Workshop which was supported by the
University of Florida and the NSF grant listed below. The authors
acknowledge valuable discussion among the workshop participants,
particularly Ruth Murray-Clay, Kaitlin Kratter, Althea Moorhead and
Andrew Youdin. The authors also thank Aaron Boley, John Johnson,
Christian Marois, Greg Laughlin and Peter Bodenheimer for input on this
work. The referee, Richard Durisen, provided particularly valuable
insight into the current state of the gravitational instability
subfield. This material is based upon work supported by the National
Science Foundation under Grant No.\ 0707203.

\clearpage
\begin{deluxetable}{cccccl}
\tabletypesize{\scriptsize}
\tablecaption{Fastest-Growing Modes in Maximum-Mass Disks \label{modes}
}
\tablehead{ \colhead{Star Type} & \colhead{$\Omega$ at Disk Edge
$(s^{-1})$} & \colhead{Pattern Speed\tablenotemark{a}}  &
\colhead{Growth Rate} & \colhead{Corotation (AU)} &
\colhead{Position in Figure 3} }
\startdata
\multirow{3}{*}{A} & \multirow{3}{*}{$2.93 \times 10^{-10}$} & 15.5 &
3.2 & 15.2 & not pictured \\
& & 2.1 & 1.1 & 59.4 & lower left \\
& & 5.1 & 0.5 & 32.3 & upper left \\
\hline
\multirow{3}{*}{G} & \multirow{3}{*}{$2.44 \times 10^{-10}$} & 3.9 & 7.1
& 38.8 & upper middle \\
& & 3.2 & 2.9 & 44.4 & not pictured \\
& & 4.9 & 2.5 & 51.6 & lower middle \\
\hline
\multirow{2}{*}{M} & \multirow{2}{*}{$1.79 \times 10^{-10}$} & 5.4 & 4.2
& 30.2 & upper right \\
& & 2.4 & 2.5 & 53.7 & lower right \\
\enddata
\tablenotetext{a}{Pattern speed and growth rate are given in units of
the orbital angular speed $\Omega$ at the outer disk edge.}
\end{deluxetable}

\clearpage

\end{document}